\documentclass[pra,twocolumn,nofootinbib,floatfix,10pt]{revtex4-2}

\usepackage{amsmath}
\usepackage{amssymb}
\usepackage{wasysym}
\usepackage{graphicx}
\usepackage{color,soul}
\usepackage{physics}
\usepackage{siunitx}
\usepackage{dsfont}
\usepackage{float}
\usepackage[english]{babel}
\usepackage{blindtext}
\usepackage[english,nomargin,inline,marginclue,draft]{fixme}
\pdfpageheight\paperheight
\pdfpagewidth\paperwidth

\usepackage[colorlinks,linkcolor=blue,anchorcolor=blue,citecolor=blue,urlcolor=blue]{hyperref}

\fxusetheme{colorsig}
\FXRegisterAuthor{cg}{acg}{CG}  
\FXRegisterAuthor{th}{ath}{\color{blue}TH}  
\FXRegisterAuthor{ib}{aib}{\color{red}IB} 
\FXRegisterAuthor{sh}{ash}{\color{cyan}SH} 
\FXRegisterAuthor{db}{adb}{\color{green}DB} 
\FXRegisterAuthor{ps}{aps}{PS}
\makeatletter
\renewcommand*\FXLayoutInline[3]{%
  {\@fxuseface{inline}\ignorespaces{\color{fx#1}[#3: #2]}}}
\makeatother

\long\def\symbolfootnote[#1]#2{\begingroup%
\def\thefootnote{\fnsymbol{footnote}}\footnotetext[#1]{#2}\endgroup}

\def\nobreakbefore{%
  \relax\ifvmode\else
    \ifhmode
      \ifdim\lastskip > 0pt\relax
        \unskip\nobreakspace
      \else 
        \nobreakspace
      \fi
    \fi
  \fi
}
\let\oldcite\cite
\renewcommand\cite{\nobreakbefore\oldcite}

\begin{document}
\title{Quantum synchronization at the critical point of Floquet driven Rydberg atoms}

\author{Bang Liu$^{1,2}$}
\author{Li-Hua Zhang$^{1,2}$}
\author{Yu Ma$^{1,2}$}
\author{Tian-Yu Han$^{1,2}$}
\author{Qi-Feng Wang$^{1,2}$}
\author{Jun Zhang$^{1,2}$}
\author{Zheng-Yuan Zhang$^{1,2}$}
\author{Shi-Yao Shao$^{1,2}$}
\author{Qing Li$^{1,2}$}
\author{Han-Chao Chen$^{1,2}$}
\author{Ya-Jun Wang$^{1,2}$}
\author{Jia-Dou Nan$^{1,2}$}
\author{Yi-Ming Yin$^{1,2}$}
\author{Guang-Can Guo$^{1,2}$}
\author{Dong-Sheng Ding$^{1,2,\textcolor{blue}{\dag}}$}
\author{Bao-Sen Shi$^{1,2}$}

\affiliation{$^1$Key Laboratory of Quantum Information, University of Science and Technology of China, Hefei, Anhui 230026, China.}
\affiliation{$^2$Synergetic Innovation Center of Quantum Information and Quantum Physics, University of Science and Technology of China, Hefei, Anhui 230026, China.}

\date{\today}

\symbolfootnote[2]{dds@ustc.edu.cn}

\begin{abstract}
The criticality enhanced correlations and susceptibility allow weak periodic driving to induce collective synchronization due to critical slowing down, providing a unique platform to study non-equilibrium order emergence. This establishes a powerful paradigm for investigating non-equilibrium order formation, yet the fundamental mechanisms of critical-point synchronization remain poorly understood. Here, we utilize a microwave pulse as a seed to induce quantum synchronization at the critical point of Floquet driven Rydberg atoms. In the experiment, the microwave periodic driving on Rydberg states acts as a seeded temporal order in subspace, which triggers synchronization across the entire ensemble. The behavior of the emergent synchronized oscillation is elaborately linked to alterations in the seed, such as the relative phase shift and the frequency difference, which result in phase-dependent seeding and embryonic synchronization. This result opens up new possibilities for studying and harnessing time-dependent quantum many-body phenomena, offering insight into the behavior of complex many-body systems under seeding.
\end{abstract}

\maketitle

Synchronization, a fundamental phenomenon in classical nonlinear systems, has been extensively studied in various contexts, from biological oscillators to engineered systems \cite{blekhman1988synchronization,pikovsky2001synchronization,boccaletti2002synchronization,Cui2024,Kim2021,traub1982cellular,Jang2018,Blasius1999}. In recent years, this concept has been extended to the quantum domain, revealing unique behaviors that differ markedly from their classical counterparts \cite{PhysRevLett.129.250601,PhysRevLett.97.210601,PhysRevLett.121.063601,PhysRevLett.112.094102,PhysRevLett.118.243602}. Quantum synchronization refers to the emergence of coordinated, phase-locked dynamics in a quantum many-body system, where microscopic degrees of freedom (e.g., spins or atomic populations) align their oscillations in time due to interconnected interactions or external perturbations \cite{PhysRevLett.100.014101,PhysRevLett.111.103605,PhysRevLett.111.234101,PhysRevLett.112.204101,PhysRevLett.123.023604,PhysRevLett.125.013601,Shi2025,OcampoEspindola2025,Clugru2020}. There has been significant progress in both theory and experiments, including studies on limit-cycle oscillators and superconducting qubits, which have highlighted the interplay between noise, dissipation, and quantum coherence in achieving synchronization \cite{PhysRevLett.88.230602,PhysRevLett.93.204103,PhysRevLett.98.184101,PhysRevLett.107.118102,PhysRevLett.121.053601,Wu2025,Vinokur2008,Kumar2025,Awad2016,Brzobohat2023}. These works not only deepen our understanding of quantum many-body dynamics but also pave the way for developing technologies in controlled quantum synchronization.

The long-range interaction offered by Rydberg atoms \cite{saffman2010quantum,adams2019rydberg,browaeys2020many} provides a good platform for studies of the non-equilibrium phase transitions \cite{ding2019Phase,ding2022enhanced,carr2013nonequilibrium}  and non-equilibrium dynamics \cite{ding2023ergodicity, wadenpfuhl2023emergence, wu2023observation} in many-body physics. The observed oscillations possess a certain degree of order and behavior that goes against the expected behavior dictated by the second law of thermodynamics, allowing us to study continuous and discrete time crystals (DTCs) and the criticality of phase transition \cite{wu2023observation,liu2024bifurcation,liu2024higher,yao2017discrete}. Near critical points, where the correlations become long-ranged and susceptibility to perturbations is enhanced, even weak periodic driving can induce collective synchronization with minimal energy cost, offering a unique opportunity to probe how non-equilibrium order emerges from microscopic interactions. However, the investigation of collective synchronization near critical points of many-body systems remains largely unexplored - this is specifically why we focus on this regime.

\begin{figure}
\centering
\includegraphics[width=1.0\columnwidth]{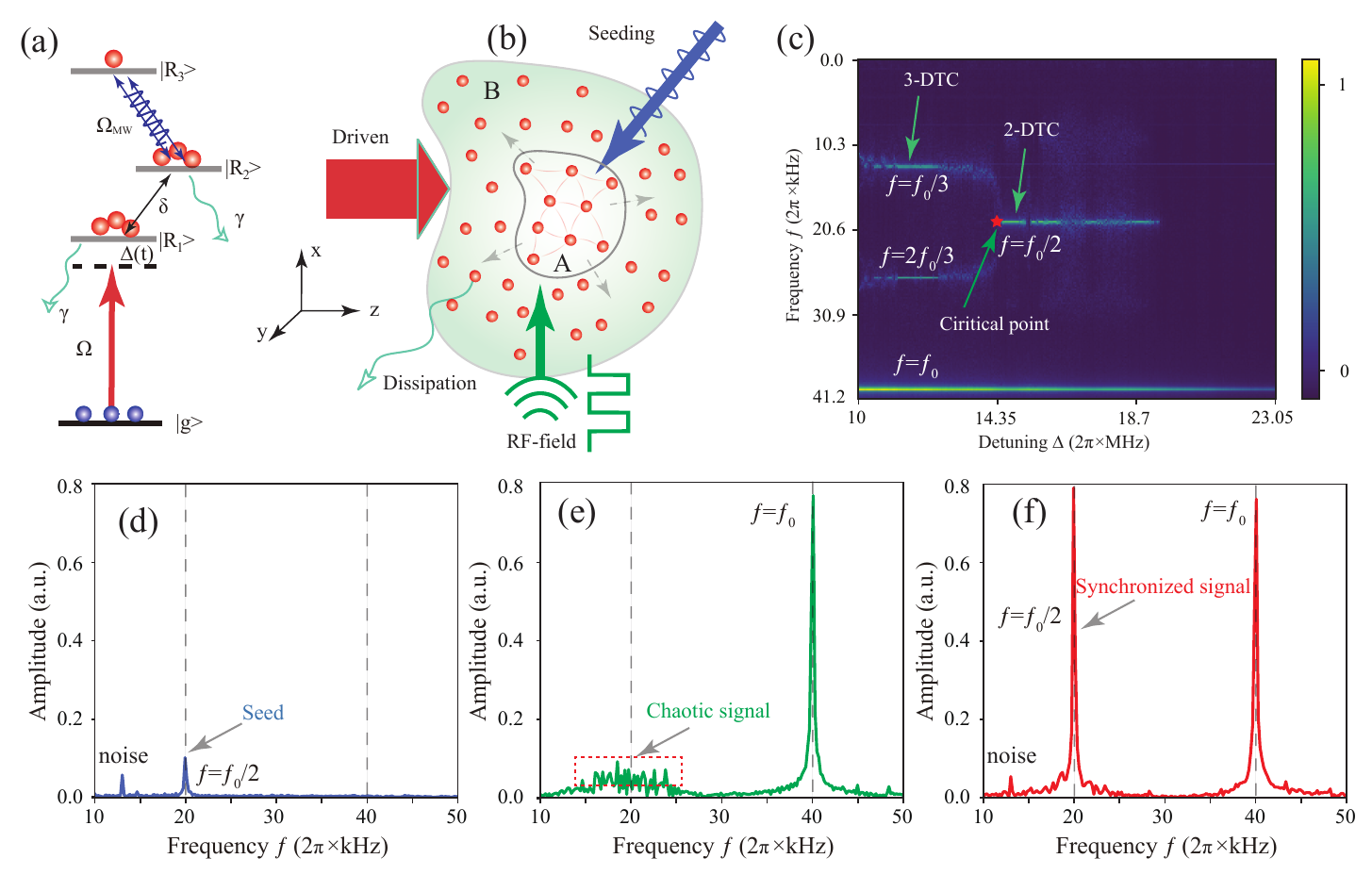}\\
\caption{\textbf{Physical model of quantum synchronization.} (a) Energy level diagram, which includes a ground state $\ket{g}$ and Rydberg states $\ket{R_{1,2}}$, the detuning $\Delta(t)$ can be shifted periodically by applying an external pulsed radio-frequency (RF) field, both of Rydberg states have a decay rate of $\gamma$. A laser drives the atom ground state $\ket{g}$ to $\ket{R_{1,2}}$ with Rabi frequency $\Omega_{1,2}$, and another amplitude modulated microwave field couples the states $\ket{R_{2}}$ and $\ket{R_3}$ with a Rabi frequency $\Omega_{\text{MW}}$. (b) Physical diagram in phase space containing many-body synchronization in a system composed of driven and dissipation Rydberg atoms and the driving RF field. (c) Measured phase maps of no-discrete time crystal (DTC), 2-DTC, 3-DTC, and their transitions. The color bar represents the Fourier transform intensity. The status of system is in a critical state before seeding which is not a discrete time crystal as shown by the red star. (d-f) The measured Fourier spectrum of input seed signal without RF field (d), Fourier spectrum with absent of seed signal (e), and Fourier spectrum with both of inputting seed signal and RF field (f). The system remains in a chaotic regime at the critical point, where the spectral response shows a finite width centered around $f_0/2$, as shown in the red dashed box in (e). The weak signal on the left side of the spectrum is the system noise.}
\label{setup}
\end{figure}

In this work, we have experimentally observed quantum synchronization at the criticality of the strongly interacting Rydberg atoms under external radio-frequency (RF) field periodic driving conditions. By applying an amplitude-modulated microwave field to periodically driving Rydberg atoms, we establish a synchronization seed that coherently locks the dynamics of the entire system. The characteristics of the amplified dynamical signal exhibit a strong dependence on the seed parameters. Our approach provides tunability over both the synchronized response frequency and phase. Through systematic variation of the seed frequency, we observe distinct dynamical responses arising from modified temporal configurations. These findings on controlled quantum synchronization open new possibilities for manipulating non-equilibrium quantum systems and engineering desired dynamical properties in driven many-body systems.

\textit{Physical Model and Experimental Diagram}\textemdash To demonstrate the physical concept of quantum synchronization at the critical point, we first build a many-body system with periodic Floquet driving, in which the system can produce discrete time crystals and the criticality \cite{liu2024higher}. The system energy diagram is depicted by Fig.~\ref{setup}(a), which contains $N$ atoms consisting of ground state $\ket{g}$ and Rydberg states $\ket{R_{1,2}}$. The Rydberg states $\ket{R_1}$ and $\ket{R_2}$ are two Rydberg states, which correspond to the sidebands induced by the RF field in the experiment. The detuning between these two Rydberg states $\delta$ corresponds to the frequency of the RF field. By applying square wave modulation to the detuning $\Delta(t) = \Delta_0 + \Delta$ when $0\leq t\text{ }\textless \text{ }T/2$, and $\Delta(t) = \Delta$ when $T/2 \leq t \text{ }\textless \text{ }T$, thus this system satisfies discrete time translation symmetry under this periodic Floquet driving. To demonstrate the process of quantum synchronization, an amplitude modulated microwave field with a modulated frequency of $f_s$ is used to drive the Rydberg states between $\ket{R_{2}}$ and $\ket{R_3}$. The physical diagram is represented in Fig.~\ref{setup}(b). Before seeding, the system exists in chaotic states that are ergodic in space B. After seeding, the system is guided into a specific non-equilibrium state, the ergodicity of the system states is broken into subspace A due to many-body interaction, with a frequency matching the seeding drive.

\begin{figure}
\centering
\includegraphics[width=1\columnwidth]{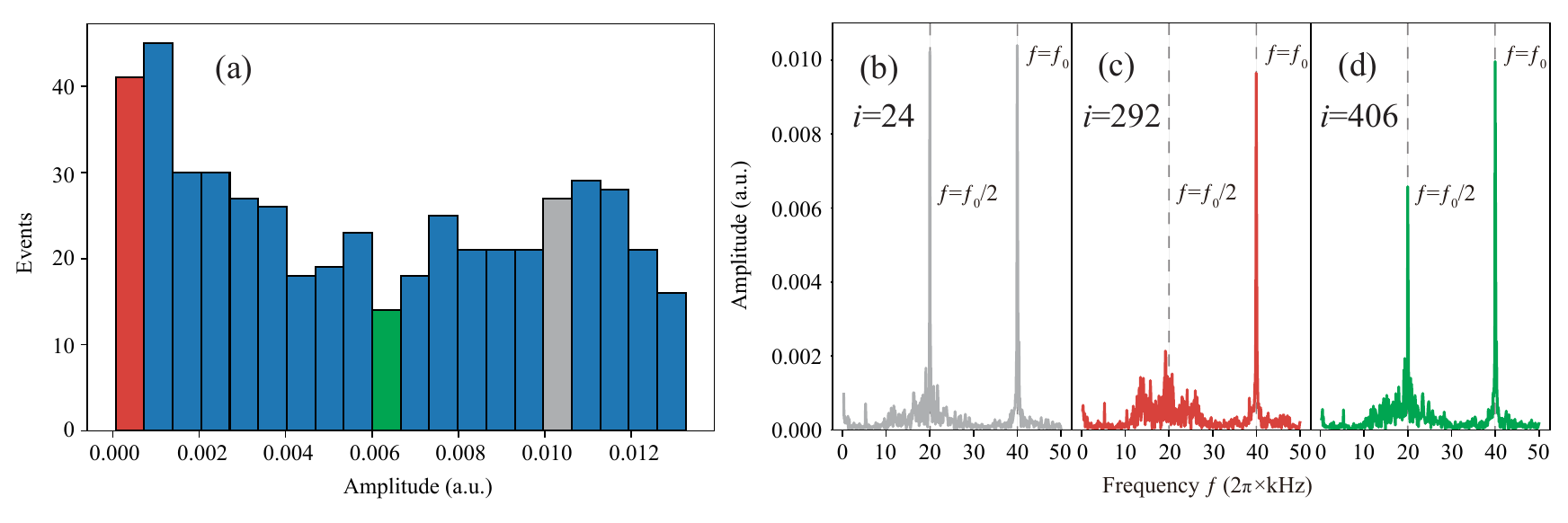}\\
\caption{\textbf{Random synchronization effect.} (a) Histogram of the amplitude of synchronized signal acquired from $i$ = 1$\sim$500 independently experimental trials. In these measurements, the relative phase between the seed field and the RF-field is fixed. (b-d) The examples for illustrating the Fourier spectrum within the corresponding regimes shown in (a) [see red, green, and gray bars]. The measured Fourier spectrum for each example corresponds to a specific experimental trial, with $i$ values of 24 (b), 292 (c), and 406 (d), respectively.}
\label{disordered}
\end{figure}

The Hamiltonian of quantum synchronization is based on periodical double Rydberg state model with additional microwave driving \cite{wu2023observation,liu2024higher}: 
\begin{equation}
\begin{aligned}
    \hat{H}(t) & =\frac{1}{2}\sum_{i}\left(\Omega_{1}\sigma_{i}^{gR_1}+\Omega_{2}\sigma_{i}^{gR_2}+\Omega_{\text{MW}}(t)\sigma_{i}^{R_2R_3}+h.c.\right)\\ &-\sum_{i}\left(\Delta(t) n_{i}^{R_1}+(\Delta(t) +\delta)n_{i}^{R_2}+(\Delta(t) +\delta)n_{i}^{R_3}\right) \\ &+\sum_{i\neq j}V_{ij}\bigg[n_{i}^{R_1}n_{j}^{R_2}+n_{i}^{R_2}n_{j}^{R_3}+n_{i}^{R_1}n_{j}^{R_3}\\ &+\frac{1}{2}(n_{i}^{R_1}n_{j}^{R_1}+n_{i}^{R_2}n_{j}^{R_2}+n_{i}^{R_3}n_{j}^{R_3})\bigg]
\end{aligned}\label{Hamiltonian}
\end{equation}
where $\Omega_{\text{MW}}(t) = A_0+A \rm{Sin}(2\pi $$f_s$$ t+\varphi_0+\varphi)$ ($f_s$ is the modulated frequency, $\varphi_0$ is the initial phase,  $\varphi$ represents the relative phase between RF-field and seed field.),  $\sigma_{i}^{gr}$ ($r={R_1,R_2}$) represents the $i$-th atom transition between the ground state $\left| g \right\rangle$ and the Rydberg state $\left|  r \right\rangle$, $n_{i}^{R_1,R_2,R_3}$ are the population operators for the Rydberg energy levels $\left|  R_1 \right\rangle$, $\left|  R_2 \right\rangle$ and $\left|  R_3 \right\rangle$, and $V_{ij}$ are the interactions between the Rydberg atoms. The Rydberg states $\left|  R_2 \right\rangle$ and $\left|  R_3 \right\rangle$ are not degenerate which are coupled resonantly by a microwave.

\begin{figure}
\centering
\includegraphics[width=1\columnwidth]{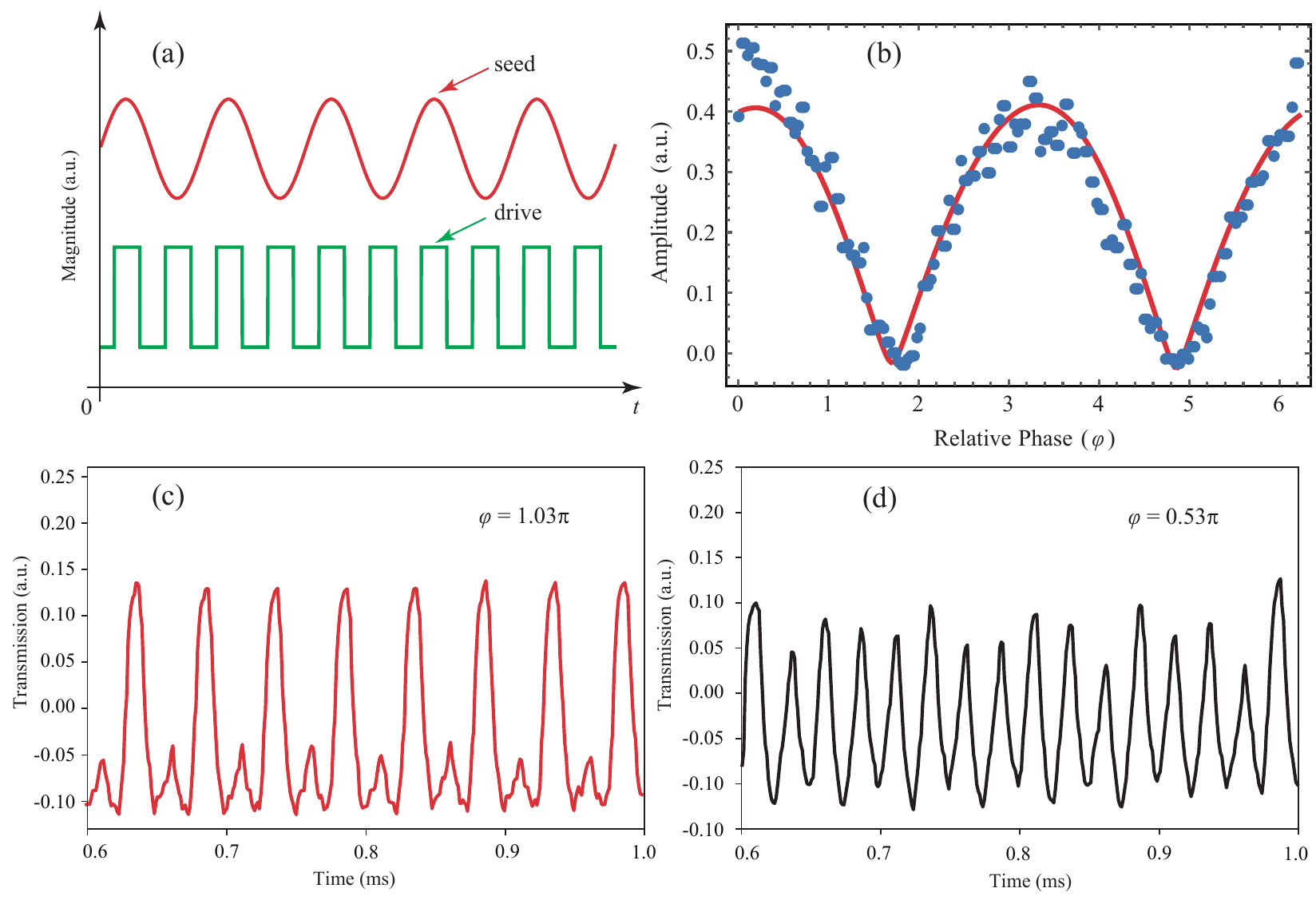}\\
\caption{\textbf{ Phase-locked synchronization.} (a) The time sequence of the driving RF-field (green) and the microwave seed field (red). The seed has a sinusoidal function and the RF-driver is a TTL signal. In this case, the periodicity of the seed is twice of the driving. The measured amplitude of synchronized oscillations (solid dots) versus the relative phase $\varphi$ between the driving field and the seed field. The red curve is the theoretical fit. (c) and (d) are the measured responses with relative phases $\varphi=1.03 \pi$ and $\varphi=0.53\pi$.}
\label{synchronization}
\end{figure}

In the experiment, we applied a three-photon electromagnetically-induced transparency (EIT) scheme to prepare the Rydberg atoms [from the ground state $\ket{6S_{1/2}}$ to the Rydberg state $\ket{49P_{3/2}}$ using three lasers with wavelengths of 852 nm (probe), 1470 nm, and 780 nm], and measured the Rydberg atom population by the transmission of the probe field \cite{zhangRydberg,liuHighly}. A microwave field couples the Rydberg states $\ket{48D_{5/2}}$ and $\ket{49P_{3/2}}$, acting as a seed. When $\Omega_{\text{MW}}=0$, the system has discrete time translation symmetry breaking with the presence of the periodic detuning $\Delta(t)$, resulting in an exotic phase diagram including a second-order DTC, higher-order DTC, and the criticality \cite{liu2024higher}, see the measured phase diagram without the seed field in Fig.~\ref{setup}(c). 

In the vicinity of the critical point, the system is sensitive to the perturbations due to the critical slowing down, and even small perturbations can have a large impact on the system's behavior. We apply a periodically modulated microwave field with the frequency of $f_s=f_0/2$ as a seed ($\Omega_{\text{MW}}\neq0$) and turn off the RF field. The population of Rydberg states is periodically modulated, so the spectrum of the system response also exhibits the $f_0/2$ frequency [see Fig.~\ref{setup}(d)]. When only the RF field is applied and the laser is tuned to near the critical point, there is no obvious subharmonics response in the Fourier spectra but there is a tendency for it to appear as the system in multiple possible non-equilibrium states near $f_0/2$ [corresponding to chaotic states], as given in Fig.~\ref{setup}(e). At the criticality, the system is in a chaotic state which exhibits response within a finite bandwidth centered around $f_0/2$. At this stage, a microwave field within this bandwidth could trigger synchronization. When seeding a weak microwave field near the critical point, a synchronized signal appears, see the results in Fig.~\ref{setup}(f). In this case, the measured ratio of output to input is greater than 7, exhibiting amplification of the seed signal. In this process, the system is guided into a specific non-equilibrium state with a frequency matching the seeding drive, the ergodicity of the system states is broken due to many-body interaction.

\textit{Random synchronization effect}\textemdash Near the critical point, the system is disordered and has a tendency to be multiple possible non-equilibrium states. When the seed field is applied, the system follows different trajectories and choose different non-equilibrium states, resulting in randomization of the phase of the atoms. By repeatedly seeding process with 500 times, we measure the amplitude of synchronized response and plot histogram of the resulting non-equilibrium states. Figure.~\ref{disordered}(a) shows the corresponding histograms of the amplitude of synchronized signal for all possible non-equilibrium states that highlight the randomization in the statistical distribution. The random counting statistics confirms the disordered phases at the criticality. 

\begin{figure}
\centering
\includegraphics[width=1.0\columnwidth]{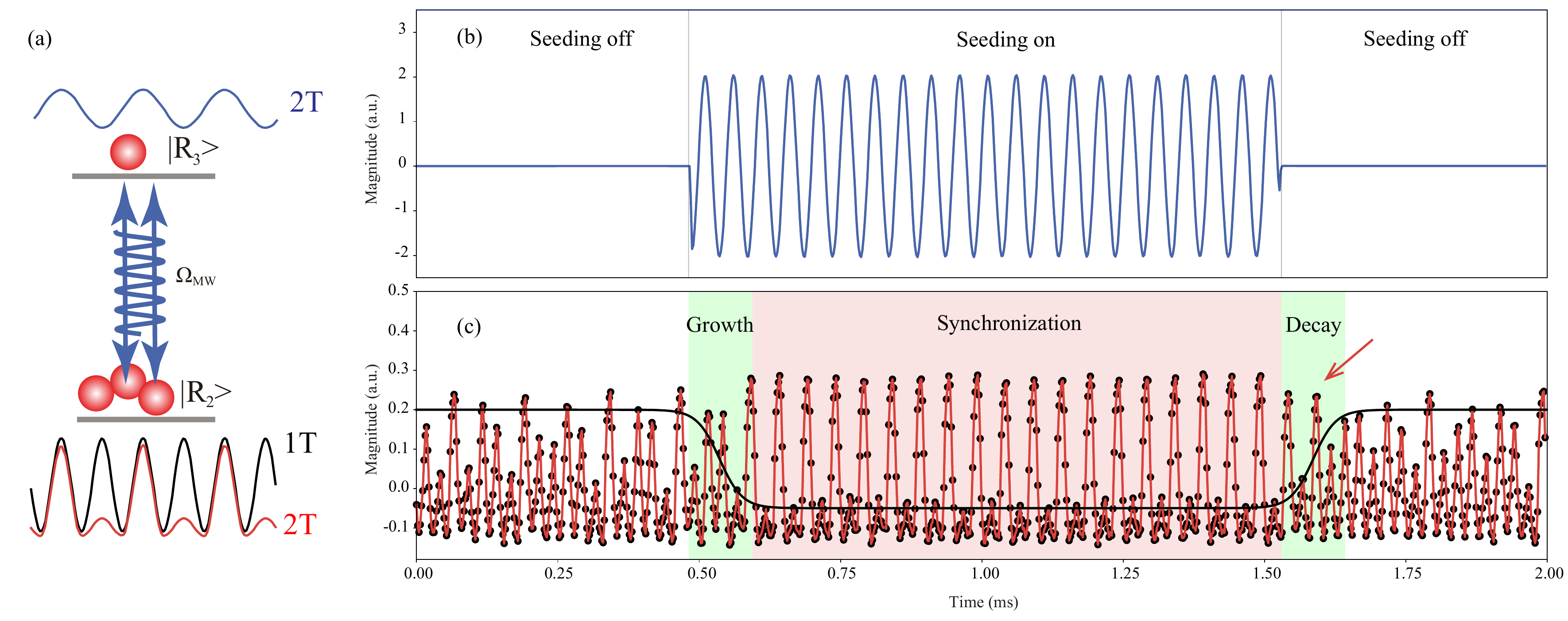}\\
\caption{\textbf{Seeding dynamics.} (a) Physical diagram for triggering synchronization.  
 The frequency of driving the Rydberg states $\ket{R_{2}}$ and $\ket{R_3}$ is twice of the frequency of the RF-field driving. The emergence of synchronized signal is triggered by switching the microwave field on. (b) The time sequence of the microwave seed field. The seeding is initiated by switching on the microwave field. (c) The recorded transmission dynamics. The transmission dynamics can be divided into three regimes: the growth regime, the synchronization regime, and the decay regime.}
\label{dynamics}
\end{figure}

We show three examples to illustrate this randomization, see the measured results presented in Figs.~\ref{disordered}(b-d). Figures.~\ref{disordered}(b-d) display the Fourier spectrum at the experimental trial $i$ = 24, 292, and 406, respectively. The amplitude of the seeded synchronization is randomized with each trial, while the phase remains preserved for an extended period after the seeding operation. These seeded non-equilibrium states occur diverse due to the fluctuations near the critical point.

\textit{Phase-Locked Seeding}\textemdash The emergence of the synchronized signal is influenced by the relative phase between the seed field and the RF-driving field. To illustrate this, we introduce a seed signal with a double periodicity ($f_s=f_0/2$) into the microwave field and vary the relative phase $\varphi$ after the seeding operation, as shown in Fig.~\ref{synchronization}(a). The seed field builds up a transition channel between Rydberg states $48\ket{D_{5/2}}$ and $49\ket{P_{3/2}}$, modifying the system states in the subspace. As the intensity of the seed field varies sinusoidally, the population at the Rydberg state $49\ket{P_{3/2}}$ also undergoes periodic evolution. This periodic decrease in population acts as a trigger for the system to synchronize, leading to the formation of a coherent and time-ordered pattern in the system.

By varying the relative phase $\varphi$ from 0 to 2$\pi$, we record the amplitude of synchronized signal, the results are given in Fig.~\ref{synchronization}(b). The changing seed waveform and the RF-driving waveform start to align with each other, causing them to overlap and leading to a joint action on exciting atoms. This joint action induces a new period of collective dynamics, see more details in supplementary materials. The response of synchronized signal versus $\varphi$ behaves as a doubling frequency with relative to the seed field. For example, when $\varphi=1.03\pi$, the synchronized signal can be amplified and we can thus observe the visible periodic doubling signal [see Fig.~\ref{synchronization}(c)]. While at $\varphi=0.53\pi$, the signal cannot be amplified, corresponding to unsynchronization. 

\begin{figure}
\centering
\includegraphics[width=1\columnwidth]{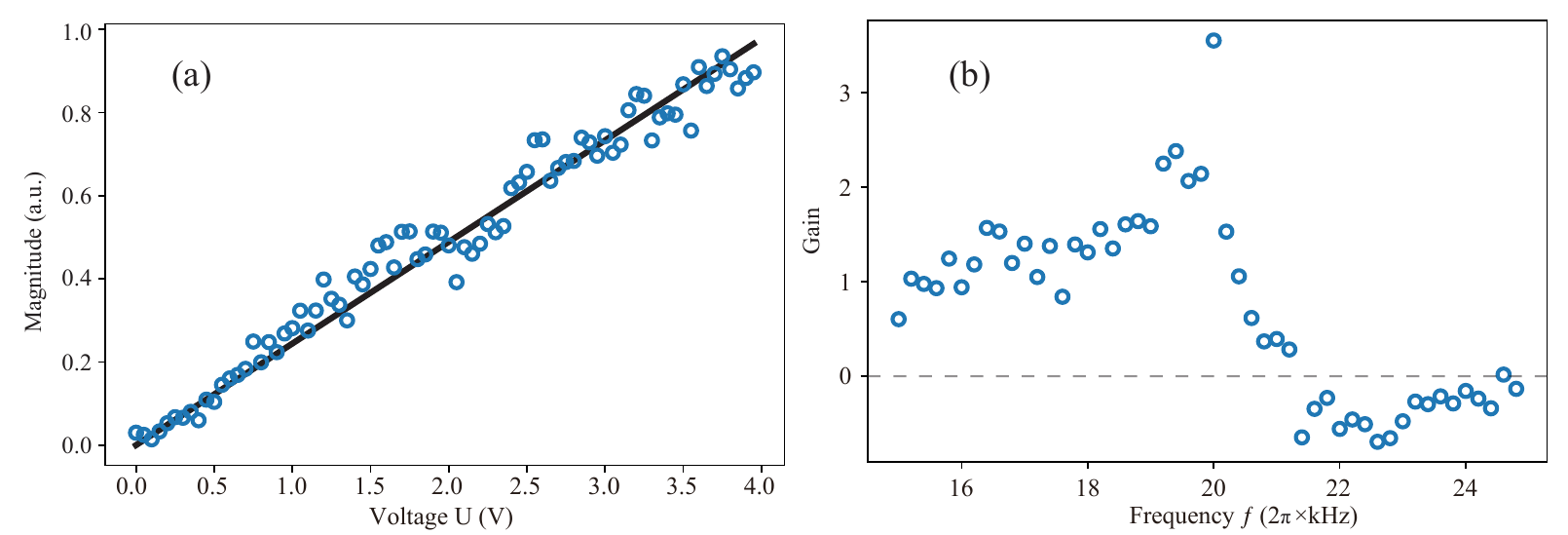}\\
\caption{\textbf{Embryonic synchronization.} (a) Synchronized response versus the seed field amplitude $U$ (blue dots). The black line is the fit linear function. Here, the seed frequency is set at $f_s = f_0/2$. (b) Measured gain by seeding with altering different $f_s$. Notably, we observe that the subharmonics align with the seed frequency. There is a higher gain at frequencies above $f_s$ and a weaker response at frequencies below $f_s$. In these cases, the relative phase $\varphi = 1.03\pi$.}
\label{frequency}
\end{figure}

\textit{Seeding Dynamics}\textemdash To demonstrate the seeding dynamics, we input a pulsed amplitude-modulated microwave field [with a frequency of $f_s = f_0/2$] to switch the seeding field off and on. The pulsed seeding field drives transition between two Rydberg states periodically, as given in Fig.~\ref{dynamics}(a). This pulsed transition process acts as a pulsed seed to affect the temporal arrangements of Rydberg population $\rho_{R_{2}R_{2}}$ in the presence of RF-field periodic driving. Specifically, it leads to the response frequency of Rydberg population $\rho_{R_{2}R_{2}}$ changing from $f_0$ to $f_0/2$ when switching the seed field on. 

We investigate the seeding dynamics by recording the probe transmission, as shown in Figs.~\ref{dynamics}(b) and (c). There are gradually modified temporal arrangements of transmission at the beginning of switching, indicating the seeding growth regime. The duration of this growth regime in our experiment is approximately $\sim$ 0.1 ms, as indicated by the shaded green area around $t$ = 0.5 ms. Once the growth regime is completed, synchronization is successfully seeded, as shown by the shaded red area in Fig.~\ref{dynamics}(c). In this regime, the system exhibits persistent periodic oscillations, but with the same frequency as the seed field. When we switch off the seed field, the synchronized signal gradually disappears, corresponding to the decay of synchronization. The system response continues to oscillate when the seed is turned off, but this persists only for a few tens of microseconds because of the system dissipation. The collisions between atoms and decoherence can disrupt the synchronization among the atoms.

\textit{Embryonic Synchronization and Amplification}\textemdash Furthermore, the shape and intensities of the initial seeding also influence the evolution of many-body system, leading to the emergence of synchronization with distinct frequencies. Firstly, we measure the magnitude of the synchronized response by altering the intensities of seed field [here, we set $f_s=f_0/2$] as shown in Fig.~\ref{frequency}(a). The magnitude of synchronized response in the Fourier spectrum is directly proportional to the amplitude of seed field $U$.

The fundamental frequency $f_0/2$ of synchronized signal is associated with the characteristic oscillation. When the driving frequency matches this resonant frequency, the system can synchronize its behavior with the external driving. When the seeding frequency deviates from $f_0/2$, the non-resonant seeding introduces perturbations or disruptions to the system, causing it to explore a broader range of possible energy states or patterns of oscillation. 

Secondly, by varying the seed frequency $f_s$ from 2$\pi\times$15 kHz to 2$\pi\times$25 kHz, we record the intensity of signal at frequency $f_s$ in the Fourier spectrum. We analyze their characteristics through the magnitude of subsequent subharmonics response in the Fourier spectrum. Importantly, we observe that the subharmonics align with the frequency of seed field, resulting in embryonic synchronization, the measured data for gain of output to input are given in Fig.~\ref{frequency}(b). The gain has a formula of $G=(S_{\text{out}}-S_{\text{in}})/S_{\text{in}}$, in which $S_{\text{in(out)}}$ represents the amplitude at $f=f_s$ in Fourier spectrum without (with) RF-field. Interestingly, we not only observe a gain greater than 0 at $f_s=f_0/2$, but we also find that this gain can occur for frequencies other than $f_0/2$. For example, when we set $f_s=2\pi\times19.6$ kHz, we still observe a gain greater than 0. For $f_s\neq f_0/2$ (corresponding to seeding with different embryos), we are essentially starting the system off with different starting points, which can lead to the emergence of synchronization with distinct characteristics. This indicates that seeded synchronization is not limited to a specific frequency, offering more flexibility in controlling and amplifying the signal. This variation in frequency allows us to modify the underlying temporal arrangements and ultimately obtain different types of oscillation formations. While at frequencies significantly different from $f_0/2$, such as $f_s=f_0/6$ or $f_s=5f_0/6$, no substantial amplification effect is observed, exhibiting a distinct difference from the synchronization behavior.

Furthermore, we observe that the gain is not symmetric at $f_s=f_0/2$. The gain is present for frequencies $f_s$ below $f_0/2$, but it disappears for frequencies $f_s$ above $f_0/2$, resulting in a loss. This asymmetric behavior can be explained from the stimulated amplification of seeding in two energy levels [$f=0$ and $f=f_0/2$] involved in the amplification process. In the case of $f_s<f_0/2$, the energy of seed field is lower than the energy difference between the two energy levels. As a result, the system is injected with energy through the RF-driving and amplify the system response, leading to a gain in the overall output. On the other hand, when $f_s>f_0/2$, the frequency of the seed field is higher than half the resonant frequency $f_0$. This means that the energy of seed field is higher than the energy difference between the two energy levels, and thus the system is hard to synchronize to the change of seed field. 

\textit{Discussions}\textemdash In summary, we have studied quantum synchronization at the critical point of Floquet driven Rydberg atoms. The coherent driving on Rydberg atoms by the microwave field forms a seed, acting as a trigger, pushing the system toward a specific non-equilibrium evolution from chaotic dynamics. Through the many-body interaction of Rydberg atoms, the laser driving and dissipation, and the seed of microwave driving, the subspaces of system are coupled to each other, resulting in collective synchronization throughout the entire ensemble.

To our best knowledge, the findings in this work reveal a first experimental demonstration of synchronization at the critical point of the many-body system. The introduced technological method opens up new avenues of engineering of non-equilibrium states, unlocking for the control and manipulation temporal patterns and subharmonic responses in quantum many-body systems. Furthermore, exploring the range of frequencies for triggering synchronization and the phase-locking mechanism help us gain a deeper understanding of the underlying mechanisms involved in their formation and stability. This achievement not only advances our fundamental understanding of quantum synchronization but also has significant implications for various applications.

\section*{Acknowledgments}
We acknowledge funding from the National Key R and D Program of China (Grant No. 2022YFA1404002), the National Natural Science Foundation of China (Grant Nos. T2495253, 61525504, and 61435011). B.L. conducted the physical experiments and developed the theoretical model. D.-S.D. conceived the idea and supported the project.

\bibliography{ref}

\end{document}